\documentclass[a4paper,twoside]{article}
\newcommand{\affil}[1]{$^{\rm #1}$}
\usepackage[authoryear]{natbib}
\bibpunct{(}{)}{;}{a}{}{,}
\usepackage{graphicx}
\usepackage{fixltx2e}
\date{} 
\title{\large\bf\flushleft Improved Neutron-Capture Element Abundances in Planetary Nebulae}
\author{\parbox{\textwidth}{\flushleft
\vspace{-0.5cm}
{\it N.\ C.\ Sterling\affil{A, I}, H.\ L.\ Dinerstein\affil{B}, S.\ Hwang\affil{B},
S.\ Redfield\affil{C}, A.\ Aguilar\affil{D}, M.\ C.\ Witthoeft\affil{A}, D.\ Esteves\affil{E}, 
A.\ L.\ D.\ Kilcoyne\affil{D}, M.\ Bautista\affil{F}, R.\ Phaneuf\affil{E}, R.\ C.\ Bilodeau \affil{D},
C.\ P.\ Ballance\affil{G}, B.\ McLaughlin\affil{H}, and P.\ H.\ Norrington\affil{H}}\\
\vspace{0.4cm}
{\small \affil{A}\,NASA Goddard Space Flight Center, Code 662, Greenbelt, MD 20771, USA}\\
{\small \affil{B}\,University of Texas, Department of Astronomy, 1 University Station, C1400, Austin,
TX 78712-0259, USA}\\
{\small \affil{C}\,Astronomy Department, Van Vleck Observatory, Wesleyan University, Middletown, CT
06459, USA} \\
{\small \affil{D}\,Advanced Light Source, Lawrence Berkeley National Laboratory, One Cyclotron Road,
MS: 6R2100, University of California, Berkeley, CA 94270, USA} \\
{\small \affil{E}\,Department of Physics, MS 220, University of Nevada, Reno, NV 89557-0058, USA} \\
{\small \affil{F}\,Virginia Polytechnic Institute \& State University, Physics Department, 
Robeson Hall (0435), Blacksburg, VA 24061, USA} \\
{\small \affil{G}\,Department of Physics, Auburn University, Auburn, AL 33333, USA} \\
{\small \affil{H}\,Centre for Atomic, Molecular and Optical Physics, School of Mathematics and Physics, 
Queen's University Belfast, The David Bates Building, 7 College Park, Belfast BT7 1NN, UK} \\
{\small \affil{I}\,NASA Postdoctoral Program Fellow; Email: Nicholas.C.Sterling@nasa.gov}}}
\begin{document}
\twocolumn[
\begin{changemargin}{.8cm}{.5cm}
\begin{minipage}{.9\textwidth}
\vspace{-1cm}
\maketitle
%
%
\small{\bf Abstract:}

Spectroscopy of planetary nebulae (PNe) provides the means to investigate \emph{s}-process enrichments of neutron(\emph{n})-capture elements that cannot be detected in asymptotic giant branch (AGB) stars.  However, accurate abundance determinations of these elements present a challenge.  Corrections for unobserved ions can be large and uncertain, since in many PNe only one ion of a given \emph{n}-capture element has been detected.  Furthermore, the atomic data governing the ionization balance of these species are not well-determined, inhibiting the derivation of accurate ionization corrections.  We present initial results of a program that addresses these challenges.  Deep high resolution optical spectroscopy of $\sim$20 PNe has been performed to detect emission lines from trans-iron species including Se, Br, Kr, Rb, and Xe.  The optical spectral region provides access to multiple ions of these elements, which reduces the magnitude and importance of uncertainties in the ionization corrections.  In addition, experimental and theoretical efforts are providing determinations of the photoionization cross-sections and recombination rate coefficients of Se, Kr, and Xe ions.  These new atomic data will make it possible to derive robust ionization corrections for these elements.  Together, our observational and atomic data results will enable \emph{n}-capture element abundances to be determined with unprecedented accuracy in ionized nebulae.


\medskip
\medskip
\end{minipage}
\end{changemargin}
]
\small

\section{Introduction}

Approximately half of the neutron(\emph{n})-capture elements (atomic number $Z>30$) in the Universe are created by slow \emph{n}-capture nucleosynthesis (the ``\emph{s}-process'').  The \emph{s}-process can occur in low- and intermediate-mass stars (1--8~M$_{\odot}$), the progenitors of planetary nebulae (PNe), during the thermally-pulsing asymptotic giant branch (AGB) phase.  Free neutrons are released by the reaction $^{13}$C($\alpha$,$n$)$^{16}$O --- or $^{22}$Ne($\alpha$,$n$)$^{25}$Mg in more massive AGB stars ($>3.5$~M$_{\odot}$) --- in the intershell region between the H- and He-burning shells.  Fe-peak nuclei experience a series of \emph{n}-captures interspersed with $\beta$-decays to transform into isotopes of heavier elements.  The enriched material is transported to the stellar envelope via convective dredge-up, and is expelled into the ambient interstellar medium by stellar winds and PN ejection \citep{bus99}.

Nebular spectroscopy uniquely reveals information about \emph{n}-capture nucleosynthesis that cannot be obtained from stellar spectra.  For example, the lightest \emph{n}-capture elements ($Z=30$--36) and noble gases are not well-studied in their sites of synthesis, due to the difficulty in detecting these species in evolved stars or supernova remnants \citep[e.g.,][]{wally95}.  In contrast, these elements are readily detected in PNe.  Spectroscopy of PNe also enables investigations of \emph{s}-process nucleosynthesis in intermediate-mass stars ($M=3.5$--8~M$_{\odot}$).  These stars are obscured by dusty, optically thick circumstellar envelopes during the AGB and post-AGB phases \citep{hab96, van03} that preclude optical or UV spectroscopy.  Hence, little is known about their contribution to \emph{s}-process nuclei in the Universe \citep{trav04}.  These stars produce Type~I PNe \citep[characterized by He and N enrichments;][]{peim78} in which \emph{n}-capture elements can be detected.

The low cosmic abundances of \emph{n}-capture elements \citep{asp05} prevented their identification in nebular spectra until \citet[][hereafter PB94]{pb94} discovered emission lines of Se, Br, Kr, Xe, and possibly other trans-iron elements in the optical spectrum of the PN NGC~7027.  Many of these identifications were later confirmed with higher resolution spectroscopy performed by \citet{zh05} and \citet{sharp07}.  PB94's analysis led \citet{din01} to identify two anonymous emission lines in the $K$~band spectra of PNe as fine-structure transitions of [Se~IV] and [Kr~III].  Shortly thereafter, \citet{ster02} identified Ge~III~$\lambda$1088.46 in absorption in the UV spectra of four PNe, and found that these objects exhibit Ge enrichments indicative of \emph{s}-process nucleosynthesis in their progenitor stars.  Together, these studies determined abundances of \emph{n}-capture elements in only 11 PNe, too small a sample to characterize \emph{s}-process enrichments in PNe as a population.

\citet[][hereafter SD08]{sd08} performed the first large-scale survey of \emph{n}-capture elements in PNe by detecting the near-infrared emission lines [Kr~III]~2.199 and/or [Se~IV] 2.287~$\mu$m in 81 of 120 Galactic PNe.  To correct for the abundances of unobserved Se and Kr ions, analytical ionization correction factors (ICFs) were derived from a large grid of photoionization models \citep{ster07}.  We derived Se and Kr abundances (or upper limits) in each object, increasing the number of PNe with known \emph{n}-capture element abundances by nearly an order of magnitude.  Se and Kr were found to be enriched by a factor of two or more in $\sim$45\% of the observed PNe, indicating efficient \emph{s}-process nucleosynthesis and convective dredge-up in their progenitor stars.  Both exhibit a positive correlation with the C/O ratio, as has been found for other \emph{n}-capture elements in AGB \citep{smith90} and post-AGB stars \citep{van03}.  Type~I PNe, which have intermediate-mass progenitors, display little or no enrichment of Se and Kr compared to other PN subclasses.  These trends have been shown to be consistent with theoretical predictions \citep{kar09}.

The $K$~band Se and Kr lines were recently detected in three nearby extragalactic PNe by \citet{din09}, providing insight to the nucleosynthesis of these elements in low-metallicity environments for the first time.  The PN Hen~2-436 in the Sagittarius dwarf spheroidal galaxy exhibits large enrichments of Se and Kr (Figure~1), comparable to the most highly enriched Galactic PNe observed by SD08, while two PNe observed in the Magellanic Clouds do not appear to be enriched.

\begin{figure}[h]
\begin{center}
\includegraphics[scale=0.43, angle=0]{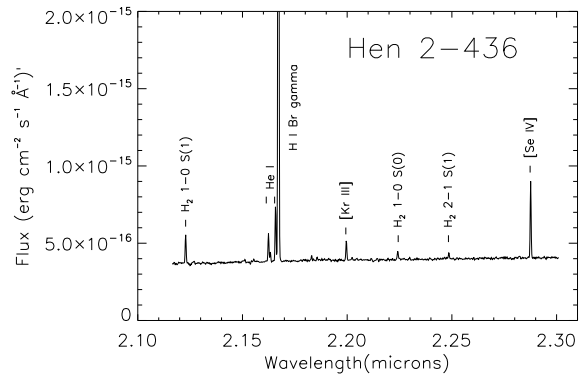}
\caption{$K$~band spectrum of the PN Hen~2-436 in the Sagittarius dwarf spheroidal galaxy, measured with the GNIRS spectrograph on Gemini South \citep[from][]{din09}.  Both [Kr~III]~2.199 and [Se~IV]~2.287~$\mu$m were detected, and are identified in the figure.}\label{fig1}
\end{center}
\end{figure}

The Se and Kr abundances derived in these studies are typically accurate to within a factor of two or three, based on the error analyses of \citet{ster07} and SD08, and similar uncertainties can be expected for other \emph{n}-capture elements observed in ionized nebulae.  These uncertainties stem from (1) the detection of only one ion each of Se and Kr, which can lead to large and uncertain ICFs; and (2) the poorly known atomic data that controls the ionization balance of these elements, which precludes the \linebreak derivation of accurate ICFs.  We present initial results from a multi-disciplinary investigation, comprised of new optical observations and atomic data determinations, designed to address these sources of uncertainty and enable much more accurate abundance analyses of \emph{n}-capture elements in ionized nebulae.

\section{Optical Observations}

Optical spectroscopy provides access to transitions from several ions of Br, Kr, Rb, Xe, and (when combined with near-infrared data) Se.  Detecting multiple ions of these elements reduces the magnitude and importance of uncertainties in their ICFs, and hence in the derived elemental abundances.  Moreover, Xe is a ``heavy-\emph{s}'' element that lies near the \emph{s}-process peak at Ba.  The abundance of Xe relative to those of lighter \emph{n}-capture elements can be used to determine the neutron exposure (i.e., the time-averaged neutron flux) experienced by Fe-peak nuclei during \emph{s}-process nucleosynthesis.  At a given metallicity, the neutron exposure controls the element-by-element distribution of \emph{s}-process enrichments \citep{bus99, bus01}.

\begin{figure*}[t!]
\begin{center}
\includegraphics[scale=0.8, angle=0]{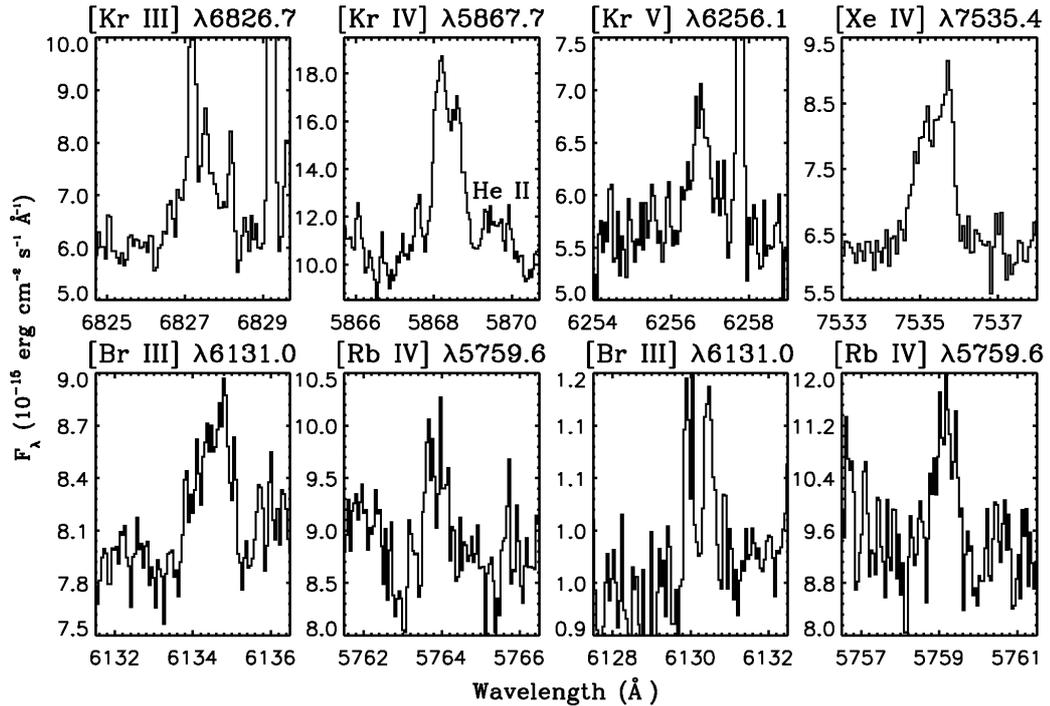}
\caption{Selected Kr and Xe emission lines detected in the PN J~900 (top panels), and Br and Rb lines in NGC~6644 (lower left panels) and NGC~3242 (lower right panels).  Telluric emission lines have not been removed, and appear as narrow features within some of the panels.  Note that wavelengths are not corrected for systemic nebular velocities.}\label{fig2}
\end{center}
\end{figure*}

The robust identification of optical \emph{n}-capture element lines in nebular spectra requires high spectral resolution in order to distinguish them from nearby transitions of more abundant elements and/or telluric features.  We are conducting optical observations of \emph{s}-process enriched Galactic PNe selected from the near-infrared sample of SD08, using the 2D~Coud\'e cross-dispersed echelle spectrometer \citep{tull95} on the 2.7-m Harlan J.\ Smith Telescope at McDonald Observatory.  Using two grating tilt settings and a 1.8''-wide slit, this instrument provides nearly complete spectral coverage in the wavelength range 3 700--10 400~\AA, at a resolution $R=36~700$ (sufficiently large to resolve many \emph{n}-capture element lines from potential blends).

To date, we have completed optical observations of approximately 17 PNe.  These objects display rich spectra, with $\sim$100 emission lines in the sparsest spectrum to over 400 lines in NGC~3242.  Kr emission has been detected in all but one of these objects, and several exhibit lines from multiple Kr ions.  At least one Xe emission line has been detected in about half of these PNe, and Se, Br, and Rb lines have been tentatively identified in 3-4 objects each.  Figure~2 displays selected Kr, Xe, Br, and Rb emission lines in targets from our sample.

\section{Atomic Data Calculations}

Abundance determinations of \emph{n}-capture elements in ionized nebulae are affected by uncertainties in the atomic data governing the ionization balance of these species --- in particular, photoionization (PI) cross-sections and rate coefficients for radiative recombination (RR), dielectronic recombination (DR), and charge transfer (CT).  These data are needed to derive accurate ICFs to correct for the abundances of unobserved ions.  Atomic data uncertainties can lead to Se and Kr abundance uncertainties of a factor of two \citep{ster07}, and similar uncertainties can be expected for other \emph{n}-capture elements.

With the atomic structure code AUTOSTRUCTURE \citep{bad86}, we are computing PI cross-sections and RR and DR rate coefficients for the first five ions of Se, Kr, and Xe, the three most widely detected \emph{n}-capture elements in PNe \citep[][SD08]{pb94, sharp07}.  This code can be used to efficiently compute multi-configuration distorted wave PI cross-sections and radiative and autoionization rates.  Relativistic effects are accounted for via the Breit-Pauli formalism and semi-relativistic radial functions.

We are computing the electronic structure of each ion using multiple configurations within the intermediate coupling scheme.  These configuration expansions are utilized to compute PI cross-sections out of the valence shells, which are resolved in both initial and final state and can be combined to determine the total cross-section of each ion.  In Figure~3, we show the computed Xe$^{3+}$ PI cross-section near the ground state threshold region, plotted against the experimental measurements of \citet{bizau06}.  RR is related to the non-resonant portion of the PI cross-sections, and can be computed from the principal of detailed balance.  Similarly, DR can be related to the resonant portion.  Our DR rate coefficients are affected by uncertainties in the energies of low-lying autoionizing levels \citep{fer03}, which have not been spectroscopically determined for these ions.  However, these uncertainties can be alleviated with our experimental PI cross-section measurements (\S4).

\begin{figure}[t]
\begin{center}
\includegraphics[scale=0.3, angle=0]{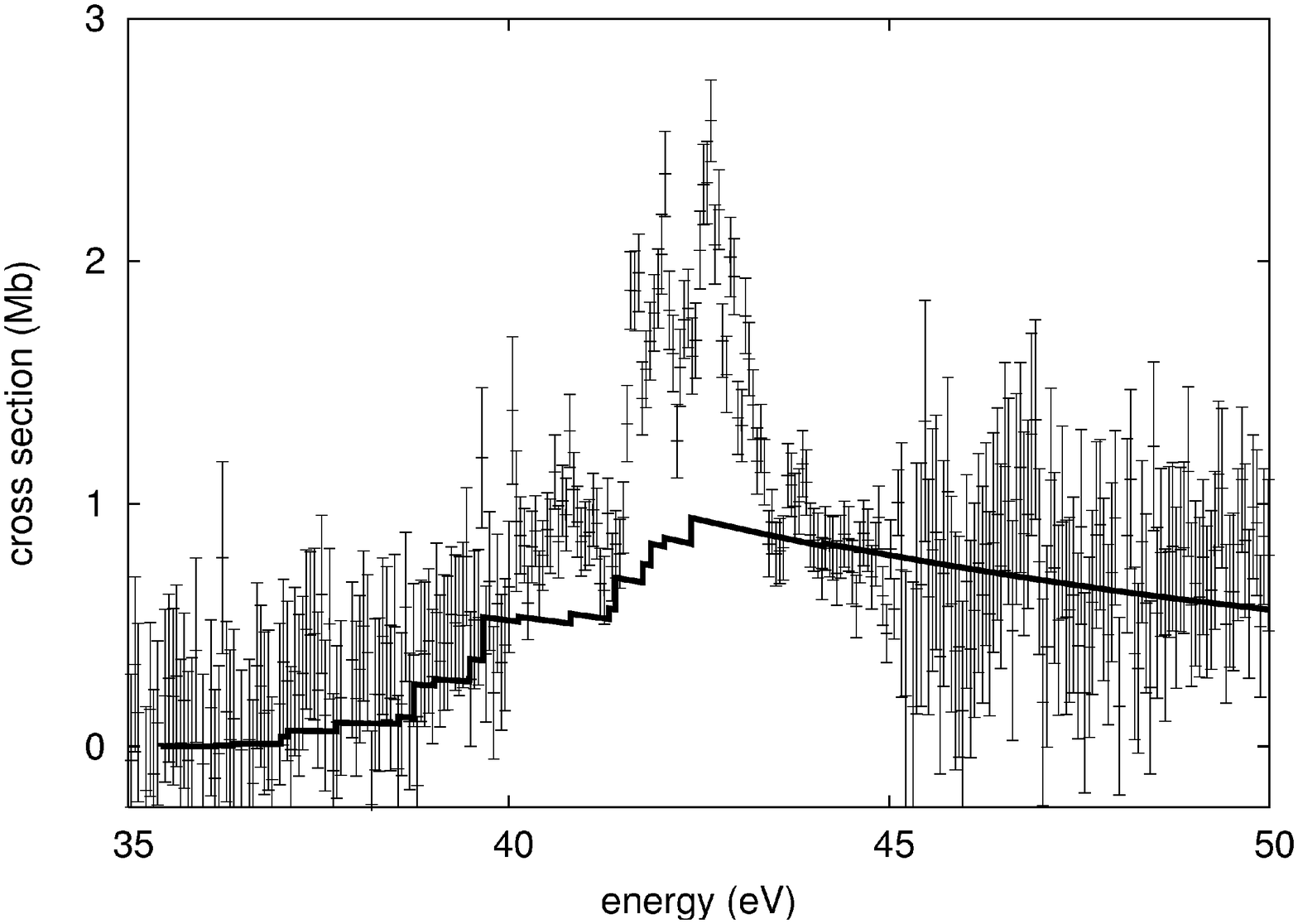}
\caption{The calculated Xe$^{3+}$ PI cross-section (solid line), using a 5 configuration expansion for Xe$^{3+}$ and 6 configurations for Xe$^{4+}$.  Note that the resonance contribution has not been added.  For comparison, we show the experimental Xe$^{3+}$ PI cross-section measured by \citet{bizau06}.}\label{fig3}
\end{center}
\end{figure}

For each of these processes, we are estimating the uncertainties in our computations in order to evaluate their effect on elemental abundance derivations (\S 5).  For example, uncertainties in the PI cross-sections and RR rate coefficients can be estimated by altering the configuration expansions utilized for the initial and target ions.  DR uncertainties are estimated by shifting the continuum level relative to the near-threshold energy levels, thereby varying the energies and number of autoionizing channels available for DR.

\section{Experimental Photoionization Cross-Section Measurements}

\begin{figure}[t!]
\begin{center}
\includegraphics[scale=0.8, angle=0]{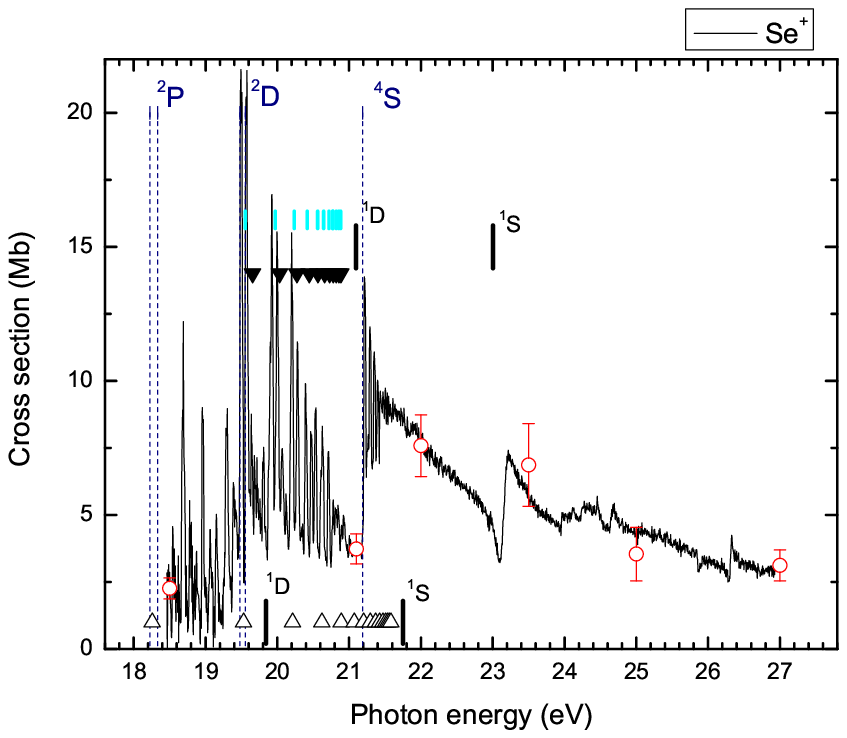}
\hfill
\includegraphics[scale=0.8, angle=0]{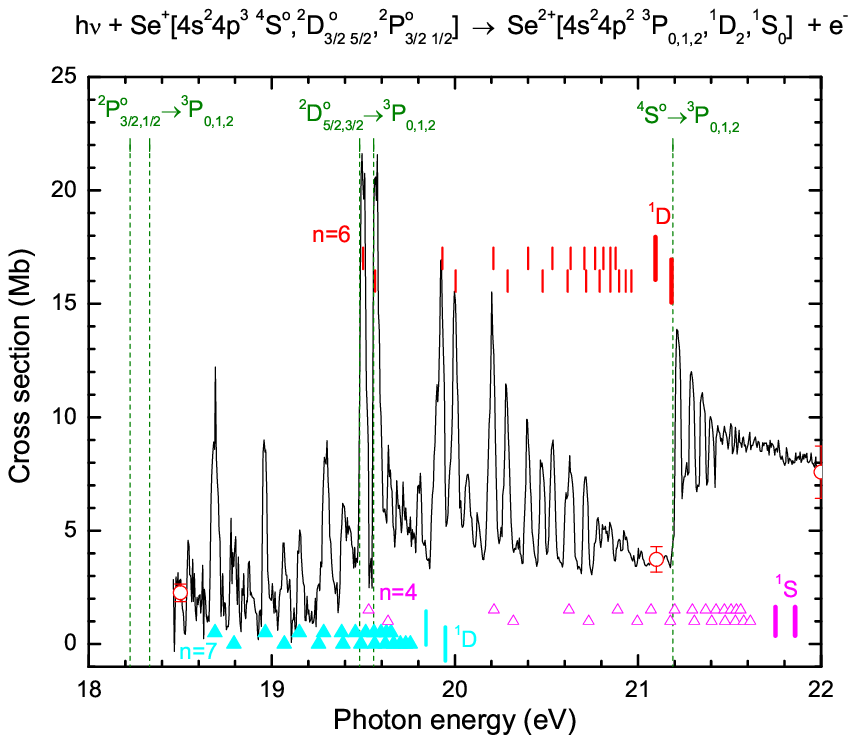}
\caption{\textit{Top panel:} The Se$^+$ photoionization cross-section measured at the ALS \citep{ster09}.  The circles with error bars represent absolute measurements performed at discrete photon energies.  \textit{Bottom panel:} Magnified view of the Se$^+$ PI cross-section in the metastable region (the ground state ionization threshold is at 21.2~eV), with resonances and Rydberg series identified.}\label{fig4}
\end{center}
\end{figure}

\begin{figure}[t]
\begin{center}
\includegraphics[scale=0.8, angle=0]{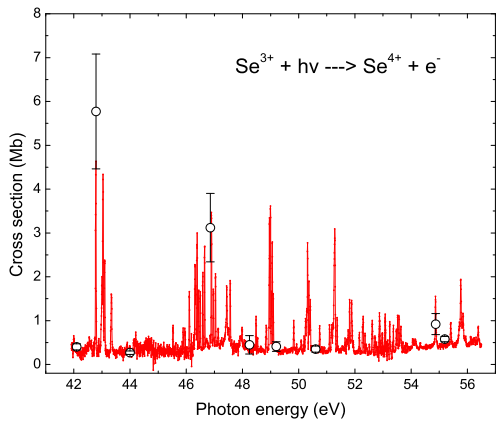}
\caption{The Se$^{3+}$ photoionization cross-section measured at the ALS.  Circles with error bars represent absolute measurements performed at discrete photon energies.}\label{fig5}
\end{center}
\end{figure}

To benchmark our atomic data calculations, we are performing experimental absolute PI cross-section measurements of Se, Kr, and Xe ions.  These data constrain both our PI and DR calculations, since PI resonances near the ionization threshold correspond to low-temperature DR resonances.  Measuring the energies of these resonances allows us to normalize the unknown energies of autoionization resonances in our DR calculations.

These measurements are being conducted at the Advanced Light Source (ALS) synchrotron radiation facility at the Lawrence Berkeley National Laboratory, with the ion-photon merged beams endstation on Beamline 10.0.1.2 \citep{cov02, ag03}.  In these measurements, a beam of atomic ions is merged with a high-energy photon beam to produce photoions.  EUV and soft X-ray photons are produced by electrons accelerated to 1.9~GeV that pass through an undulator located in a straight section of the storage ring lattice. At the beamline, the photon beam energy and resolution are controlled by a spherical-grating monochromator with three interchangeable gratings and adjustable slits.  Atomic ions are produced in an electron-cyclotron resonance source, and accelerated by a potential (typically 6~keV).  The desired charge state is selected with an analyzing magnet, and the ion beam is merged with the counter-propagating photon beam.  Photoions are produced along the merged beam path, and are directed to a detector by a demerging magnet.

This apparatus has two operational modes: spectroscopy and absolute mode.  For spectroscopy, photoion yields produced along the entire merged path are recorded in arbitrary units as a function of photon energy.  To place these yields in physical units (absolute mode), a potential is placed on a cylindrical ``interaction region,'' which has a precisely defined length.  This potential energy-tags all photoions produced within the interaction region, and the demerging magnet setting is altered so that only the energy-tagged ions are detected.  By determining the overlap of the photon and ion beams in the interaction region, absolute PI cross-sections can be measured at discrete energies.  These measurements are used to normalize the photoion yields onto an absolute scale.

We have measured absolute PI cross-sections near the ionization thresholds of Se$^+$ \citep[][Figure~4]{ster09}, Se$^{2+}$, Se$^{3+}$ (Figure~5), Xe$^{2+}$, and the photoion yield of Xe$^+$.  PI cross-sections of Kr ions have already been measured \citep{lu06a, lu06b, luth}, as have those of more highly charged Xe ions \citep{bizau06}.  Unfortunately, the Se$^{4+}$ PI cross-section cannot be measured with this apparatus due to an Auger resonance effect (also seen in the isoelectronic Kr$^{6+}$).

The analysis of these measurements is complicated by the presence of meta-stable states in the primary ion beam, which may contribute to the measured cross-sections along with ground state ions.  It is necessary to measure the PI cross-sections at energies below the ground state threshold (i.e., down to those of the metastable states) in order to estimate the fractional content of these states in the ion beam.

The measured cross-sections are interpreted with the aid of the fully relativistic $R$-matrix code DARC\footnote{http://www.am.qub.ac.uk/DARC} \citep{nor87, nor91} in order to identify resonance features and to disentangle the contributions of the ground and metastable states (e.g., Figure~6).  A parallel version of the DARC suite of codes has been developed by \citet{bg06}, and was recently modified to efficiently compute numerous bound-free dipole matrix elements, distributed over an equivalent number of processors to enable photoionization calculations to be carried out at a high degree of accuracy \citep{bal08}.

\begin{figure*}[t]
\begin{center}
\includegraphics[scale=0.4, angle=0]{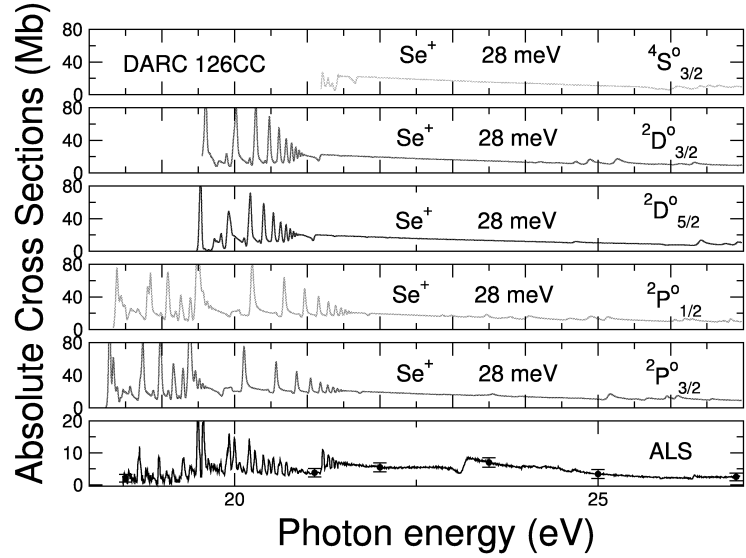}
\caption{Theoretical Se$^+$PI cross-sections (upper panels) computed from 126-level close-coupling DARC calculations \citep{ster09}, convoluted with a Gaussian to the experimental FWHM resolution of 28 meV.  The cross-section of each state in the ground configuration is shown, and compared to the experimental cross-section (bottom panel).}\label{fig6}
\end{center}
\end{figure*}

\section{Implementation}

The new atomic data will be incorporated into the atomic database of the photoionization code Cloudy \citep{fer98}.  Following the methodology of \citet{ster07}, we will compute a grid of models to determine analytical ICFs for Se, Kr, and Xe. In addition, we will use Cloudy to test the effects that our estimated atomic data uncertainties have on the derived elemental abundances.  This quantitative analysis will clarify the ionic systems and atomic processes that require further attention in future investigations.  The ICFs will be applied to our optical (\S2) and near-infrared spectra (SD08) to derive accurate \emph{n}-capture element abundances in PNe, in order to study the details of \emph{s}-process nucleosynthesis and convective mixing in PN progenitor stars.  Combined, our optical observations and atomic data determinations will allow us to determine \emph{n}-capture element abundances and \emph{s}-process enrichment factors in ionized nebulae with unprecedented accuracy.

\section*{Acknowledgments} We gratefully acknowledge J.\ Bizau, who sent us experimental PI cross-section data for Xe$^{3+}$ through Xe$^{6+}$.  NCS is supported by an appointment to the NASA Postdoctoral Program, administered by Oak Ridge Associated Universities through a contract with NASA.  BMM acknowledges support by the US National Science Foundation through a grant to ITAMP at the Harvard-Smithsonian Center for Astrophysics.  CPB is supported by US DOE grants DE-FG02-05R54819 and DE-FG02-99ER54367.  The DARC computations were conducted at the National Energy Research Scientific Computing Center in Oakland, CA.  This work has been partially supported by NASA grant 06-APRA206-0049 and NSF grants AST 0708245 and 0406809.



\begin{thebibliography}{}

\bibitem[Aguilar et al.\ (2003)]{ag03} Aguilar, A., et al.\ 2003, ApJS, 146, 467

\bibitem[Asplund et al.\ (2005)]{asp05} Asplund, M., Grevesse, N., \& Sauval, A.\ J.\ 2005, in Cosmic Abundances as Records of Stellar Evolution and Nucleosynthesis, ASP Conf.\ Ser.\ 336, eds.\ T.\ G.\ Barnes~III and F.\ N.\ Bash (San~Francisco: ASP), 25

\bibitem[Badnell (1986)]{bad86} Badnell, N.\ R.\ 1986, J.\ Phys.\ B, 19, 3827

\bibitem[Ballance \& Griffin (2006)]{bg06} Ballance, C.\ P., \& Griffin, D.\ C.\ 2006, J.\ Phys.\ B, 39, 3617

\bibitem[Ballance et al.\ (2008)]{bal08} Ballance, C.\ P., Norrington, P.\ H., \& McLaughlin, B.\ M.\ 2008, J.\ Phys.\ B, submitted

\bibitem[Bizau et al.\ (2006)]{bizau06} Bizau, J.\ M., Blancard, C., Cubaynes, D., Folkmann, F., Champeaux, J. P., Lemaire, J. L., Wuilleumier, F. J.\ 2006, Phys.\ Rev.\ A, 73, 022718

\bibitem[Busso et al.\ (1999)]{bus99} Busso, M., Gallino, R., \& Wasserburg, G.\ J.\ 1999, ARA\&A, 37, 239

\bibitem[Busso et al.\ (2001)]{bus01} Busso, M., Gallino, R., Lambert, D.\ L., Travaglio, C., \& Smith, V.\ V.\ 2001, ApJ, 557, 802

\bibitem[Covington et al.\ (2002)]{cov02} Covington, A.\ M., et al.\ 2002, Phys.\ Rev.\ A, 66, 062710

\bibitem[Dinerstein (2001)]{din01} Dinerstein, H.\ L.\ 2001, ApJ, 550, L223

\bibitem[Dinerstein et al.\ (2009)]{din09} Dinerstein, H.\ L., Sterling, N.\ C., Geballe, T.\ R., \& Wood, J.\ L.\ 2009, ApJ, in preparation

\bibitem[Ferland et al.\ (1998)]{fer98} Ferland, G.\ J., et al.\ 1998, PASP, 110, 761

\bibitem[Ferland (2003)]{fer03} Ferland, G.\ J.\ 2003, ARA\&A, 41, 517

\bibitem[Habing (1996)]{hab96} Habing, H.\ J.\ 1996, A\&ARv, 7, 97

\bibitem[Karakas et al.\ (2009)]{kar09} Karakas, A.\ I., van~Raai, M.\ A., Lugaro, M., Sterling, N.\ C., \& Dinerstein, H.\ L.\ 2009, ApJ, 690, 1130

\bibitem[Lu et al.\ (2006a)]{lu06a} Lu, M., et al.\ 2006a, Phys.\ Rev.\ A, 74, 012703

\bibitem[Lu et al.\ (2006b)]{lu06b} Lu, M., et al.\ 2006b, Phys.\ Rev.\ A, 74, 062701

\bibitem[Lu (2006)]{luth} Lu, M.\ 2006, Ph.D.\ Thesis, University of Nevada-Reno

\bibitem[Norrington \& Grant (1987)]{nor87} Norrington, P.\ H., \& Grant, I.\ P.\ 1987, J.\ Phys.\ B, 20, 4869

\bibitem[Wijesundera et al.\ (1991)]{nor91} Wijesundera, W.\ P., Parpia, F.\ A., Grant, I.\ P., \& Norrington, P.\ H.\ 1991, J.\ Phys.\ B, 24, 1803

\bibitem[Peimbert (1978)]{peim78} Peimbert, M.\ 1978, in Planetary Nebulae, IAU Symp.\  76, ed.\ Y.\ Terzian, (Dordrecht: Reidel), 215

\bibitem[P\'{e}quignot \& Baluteau (1994)]{pb94} P\'{e}quignot, D., \& Baluteau, J.-P.\ 1994, A\&A, 283, 593

\bibitem[Sharpee et al.\ (2007)]{sharp07} Sharpee, B., et al.\ 2007, ApJ, 659, 1265

\bibitem[Smith \& Lambert (1990)]{smith90} Smith, V.\ V., \& Lambert, D.\ L.\ 1990, ApJS, 72, 387

\bibitem[Sterling et al.\ (2002)]{ster02} Sterling, N.\ C., Dinerstein, H.\ L., \& Bowers, C.\ W.\ 2002, ApJ, 578, L55

\bibitem[Sterling et al.\ (2007)]{ster07} Sterling, N.\ C., Dinerstein, H.\ L., \& Kallman, T.\ R.\ 2007, ApJS, 169, 37

\bibitem[Sterling \& Dinerstein (2008)]{sd08} Sterling, N.\ C., \& Dinerstein, H.\ L.\ 2008, ApJS, 174, 158 (SD08)

\bibitem[Sterling et al.\ (2009)]{ster09} Sterling, N.\ C., et al.\ 2009, J.\ Phys.\ B, in preparation

\bibitem[Travaglio et al.\ (2004)]{trav04} Travaglio, C., Gallino, R., Arnone, E., Cowan, J., Jordan, F., \& Sneden, C.\ 2004, ApJ, 601, 864

\bibitem[Tull et al.\ (1995)]{tull95} Tull, R.\ G., MacQueen, P.\ J., Sneden, C., Lambert, D.\ L.\ 1995, PASP, 107, 251

\bibitem[Van~Winckel (2003)]{van03} Van Winckel, H.\ 2003, ARA\&A, 41, 391

\bibitem[Wallerstein et al.\ (1995)]{wally95} Wallerstein, G., Vanture, A.\ D., Jenkins, E.\ B., \& Fuller, G.\ M.\ 1995, ApJ, 449, 688

\bibitem[Zhang et al.\ (2005)]{zh05} Zhang, Y., Liu, X.-W., Luo, S.-G., P\'{e}quignot, D., \& Barlow, M.\ J.\ 2005, A\&A, 442, 249

\end{thebibliography}
\end{document}